\documentclass[11pt]{article}

\usepackage[utf8]{inputenc}
\usepackage{amsmath}
\usepackage{amsfonts}
\usepackage{amssymb}
\usepackage{float}

\newcommand{\lc}[1]{\varepsilon^{#1}} 
\newcommand{\dir}[2]{\delta^{(3)}(\vec{#1}-\vec #2)} 
\newcommand{\kr}[2]{\delta^{#1}_{#2}}
\newcommand{\U}[3]{#1^{#2}_{\phantom{#2}#3}} 
\newcommand{\vD}[2]{#1_{#2}} 
\newcommand{\MD}[3]{#1_{#2}^{\phantom{#2}#3}} 
 
\newcommand{\pb}[2]{\left\{#1,{#2}\right\}}

\let\oldnabla\nabla
\renewcommand{\nabla}{\mathbf{\oldnabla}}

\def\et{\eta}
\def\D{\Delta}
\def\d{\delta}
\def\L{\Lambda}
\def\l{\lambda}
\def\S{\Sigma}
\def\G{\Gamma}
\def\g{\gamma}
\def\e{\epsilon}
\def\s{\sigma}
\def\o{\omega}

\def\i{\iota}
\def\a{\alpha}
\def\b{\beta}

\def\m{\mu}
\def\n{\nu}
\def\r{\rho}
\def\s{\sigma}
\def\p{\pi}
\def\P{\Pi}
\def\f{\phi}
\def\F{\Phi}
\def\th{\theta}

\def\e{\epsilon}
\def\pa{\partial}

\newcommand{\be}{\begin{equation}}
\newcommand{\ee}{\end{equation}}
\newcommand{\bea}{\begin{eqnarray}}
\newcommand{\eea}{\end{eqnarray}}

\begin{document}

\begin{center}
\bf{\Large Canonical formulation of Poincar\'e BFCG theory and its quantization}
\end{center}

\bigskip
\begin{center}
 Aleksandar Mikovi\'c$\,^{a,b}$ and Miguel A. Oliveira$\,^{b,}$\footnote{E-mails: amikovic@ulusofona.pt;\ masm.oliveira@gmail.com }\\
\bigskip
{\it $\,^{a}$Departamento de Matem\'atica  \\
Universidade Lus\'ofona de Humanidades e Tecnologias\\
Av. do Campo Grande, 376, 1749-024 Lisboa, Portugal}\\
\bigskip
{\it $\,^{b}$Grupo de Fisica Matem\'atica da Universidade de Lisboa\\
Av. Prof. Gama Pinto, 2, 1649-003 Lisboa, Portugal}\\
\end{center}


\bigskip
\centerline{\bf Abstract}
\begin{quotation}
\noindent\small{We find the canonical formulation of the Poincar\'e BFCG theory in terms of the spatial 2-connection and its canonically conjugate momenta. We show that the Poincar\'e BFCG action is dynamically equivalent to the BF action for the Poincar\'e group and we find the canonical transformation relating the two. We study
the canonical quantization of the Poincar\'e BFCG theory by passing to the Poincar\'e-connection basis. The quantization in the 2-connection basis can be then achieved by performing a Fourier transform. We also briefly discuss how to approach the problem of constructing a basis of spin-foam states, which are the categorical generalization of the spin-network states from Loop Quantum Gravity.}\end{quotation}

\bigskip
\section{Introduction}

Canonical formulation of General Relativity (GR) is suitable for performing a non-perturbative and background-metric independent 
quantization of GR, see \cite{i,r}. When using the spatial metric and its canonically conjugate momentum as the degrees of freedom 
for the gravitational field, one obtains a non-polynomial Hamiltonian Constraint (HC). Consequently the corresponding operator 
in the canonical quantization yields the Wheeler-DeWitt (WdW) equation, which is difficult to solve.
 
The situation improves if the Ashtekar variables are used \cite{a}. These are given by an $SU(2)$ complex connection on the 
spatial manifold and its canonically conjugate momentum. One then obtains a polynomial HC, but since the connection is complex, 
this introduces an additional non-polynomial constraint, the reality condition, which makes the quantization complicated. 
One can also use the real Ashtekar connection \cite{br}, but then the HC becomes again non-polynomial. Still, the fact that the 
basic canonical variables are the same as the $SU(2)$ Yang-Mills gauge theory canonical variables, makes it possible to use the holonomy and the electric-field flux variables, which leads to spin-network variables and Loop Quantum Gravity, (LQG) see \cite{r}. 

The difficulties of solving the HC in the canonical LQG have led to the development of a path-integral quantization approach known as spin-foam models (SF), see \cite{bz, p}. Although the HC problem can be solved in the SF approach by using the path-integral construction of the evolution operator, there is the problem of the classical limit of a SF model \cite{sc} and the problem of the coupling of fermionic matter \cite{2pgr,sc}. These problems are related to the fact that the edge-lengths, or the tetrads, are not always defined in a spin-foam model of quantum gravity. 

Namely, the large-spin asymptotics of the EPRL-FK spin-foam model\footnote{This is the best spin-foam model of quantum gravity constructed so far, see \cite{eprl,fk}.} amplitude is a sum of $\exp(\pm i S_{\s}(j,\theta))$ terms, where $S_{\s}(j,\theta)$ is the area-Regge action for a 4-simplex $\s$, $j=(j_1,...,j_{10})$ are areas of the triangles in $\s$ and $\theta =(\theta_1,...,\theta_{10})$ are the corresponding dehidral angles, see \cite{sca}. This result can be used to calculate the effective action for the whole triangulation \cite{easf}. It is easy to see that the classical limit of the effective action is the area-Regge action. Furthermore, it was argued in \cite{easf} that the constraints which convert the area-Regge action into the Regge action were present in the effective action. However, it was showed in \cite{sc} that the effective action does not contain the Regge constraints, and hence the classical limit is just the area-Regge action. Although the area-Regge action reduces to the Regge action when the triangle areas correspond to some assigment of the edge lenghts, an arbitrary choice of the triangle areas may not correspond to any choice of the edge lengths. Hence a spin-foam geometry is not equivalent to a metric geometry, see also \cite{fs,dr}. Since the lengths of the edges in a spin foam are not always defined, then it is not possible to couple fermions, because the discrete fermionic action can be formulated only if the lengths of all the edges are defined. 

In order to introduce the edge lengths in the SF formalism, one has to introduce the tetrads in the BF theory formulation of GR. This can be done by using a formulation of GR based on the Poincar\'e 2-group \cite{2pgr}. The idea is to reformulate GR as a constrained topological theory of the BFCG type \cite{2bf}. This approach is a categorical generalization of the constrained BF-theory formulation of GR which is used for the SF models, see \cite{f}. 

The BFCG reformulation of GR is useful for the path-integral quantization. In this case one obtains the spin-cube models, which represent a categorical generalization of the SF models \cite{sc}. As far as the canonical quantization (CQ) is concerned, the progress has been hindered because the constrained BFCG theory has a complicated canonical structure. A reasonable strategy is to study first 
a simpler theory, which is the unconstrained BFCG theory. This is a topological gravity theory, and we will show that its canonical formulation is simple to understand. Another feature of this theory is that it is equivalent to the Poincar\'e group BF theory, so that one can perform a canonical quantization in terms of the BF theory variables. This is mathematically simpler than 
performing a canonical quantization in terms of the BFCG theory variables and it can also help to understand the quantization based on a spin-foam basis, which is a categorical generalization of the spin-network basis from LQG. 

In section 2 we review the Poincar\'e BFCG theory and its relation to GR. In section 3 we perform a canonical analysis of the BFCG theory by using a shortened Dirac procedure. In section 4 we reformulate the BFCG theory as a BF theory for the Poincar\'e group and find the canonical transformation which relates the two canonical formulations. In section 5 we study the canonical quantization of the BFCG theory and by using the canonical transformation from the previous section we
find a relation between the 2-connection basis and the Poincare-connection basis. We also indicate how to construct the spin-network and the spin-foam wavefunctions. In section 6 we present our conclusions.

\section{Poincar\'e BFCG theory}

Poincar\'e BFCG theory is a theory of flat 2-connections for a Poincar\'e 2-group, see \cite{2bf,2pgr}. A 2-group is a 2-category with one object where
all the 1-morphisms and all the 2-morphisms are invertible. This is equivalent
to having a pair of groups $(G,H)$ with a group action $\triangleright: G\times H \to H$ and a homomorphism $\partial : H \to G$.
The morphisms are the elements of $G$, while the 2-morphisms are the elements of the 
semi-direct product group $G\times_s H$. In the Poincar\'e 2-group case $G=SO(1,3)$ and $H={\bf R}^4$, while the group action is
given by a Lorentz transformation of a four vector from ${\bf R}^4$ and $\partial$ is trivial. The 2-morphisms form the 
Poincar\'e group $ISO(1,3)$. 

One can define a notion of a 2-connection for a Lie 2-group, in analogy to the connection on a principal bundle for a manifold $M$ 
and a Lie group $G$. The 2-connection is a pair $(A,\b)$, where $A$ is a one-form taking values in the Lie algebra $\bf g$ of $G$,
 while $\b$ is a 2-form taking values in the Lie algebra $\bf h$ of $H$. The gauge transformations of $(A,\b)$ are given by the 
usual gauge transformations
\be A \to g^{-1}(A + d)\,g \,,\quad \b \to g^{-1} \triangleright \b \,,\ee
where $g:M \to G$. These transformations correspond to local 1-morphisms, while the 2-morphisms from $H$ generate a new gauge transformation
\be A\to A \,, \quad \b \to \b + d \e + A \wedge^\triangleright \e \,,\ee
where $\e$ is a one-form from $\bf h$ and 
\be A \wedge^\triangleright \e = A^I \wedge \e^\a \,\D_{I\a}^\b T_\b \,.\ee 
Here $\D$ are the structure constants defined by the group action $\triangleright$ for the corresponding Lie algebras. Hence $X_I \triangleright T_\a = \D_{I\a}^\b T_\b$, where $X$ is a basis for $\bf g$ and $T$ is a basis for $\bf h$.

In the Poincar\'e 2-group case we have 
\be A(x)=\o^{ab}(x)\,J_{ab} \,,\quad \b(x) =\b^a (x)\, P_a \,,\ee
where $J$ are the Lorentz group generators and $P$ are the translation generators. We then obtain for 
the infinitesimal gauge transformations
\be \d_\l \o^{ab} = d\l^{ab} + \o^{[a}_c \,\l^{b]c} \,,\quad \d_\l \b^a = \l^a_c \,\b^c \,,\ee
while for the infinitesimal 2-morphism gauge transformations we obtain 
\be \d_\e \o = 0 \,,\quad \d_\e \b^a = d\e^a + \o^a_c \wedge \e^c \,.\ee

The curvature for a 2-connection $(A,\b)$ is a pair of a 2-form ${\cal F}\in {\bf g}$ and a 3-form ${\cal G}\in {\bf h}$, 
given by
\be {\cal F} = dA + A\wedge A - \partial\b \,,\quad
{\cal G} = d \beta + A \wedge^\triangleright \beta \,.\ee
In the Poincar\'e 2-group case, we have
\begin{eqnarray}
 	{\cal F}^{ab}\equiv	R^{ab}&=&d\omega^{ab}+\omega^{a}_{\phantom a c}\wedge\omega^{cb}\\
 	{\cal G}^a \equiv	G^a&=&\nabla \beta^a=d\beta^{a}+\omega^{a}_{\phantom a c}\wedge\beta^{c}\,,
 		\end{eqnarray}
so that $R^{ab}$ is the usual spin-connection curvature. The $\partial\b$ term does not appear in $R^{ab}$ beacuse $\partial\b =0$ for the Poincar\'e 2-group.	

The dynamics of flat 2-connections for the Poincar\'e 2-group is given by the BFCG action
	\begin{eqnarray}\label{BFCG}
 		S =\int_M \left( B_{ab}\wedge R^{ab}+e_{a}\wedge G^{a}\right) \label{2p}
 	\end{eqnarray}
 where $B^{ab}$ is a 2-form and $e_a$ are the tetrads \cite{2pgr}. The Lagrange multipliers $B$ and $e$ transform under the usual 
gauge transformations as
 \be  B \to g^{-1}\, B g \,,\quad e \to g\triangleright e \,, \ee
while the 2-morphism transformations are given by
\be B_{ab} \to B_{ab} + e_{[a} \wedge \e_{b]} \,,\quad e_a \to e_a \,, \ee
see \cite{2bf}. The action (\ref{2p}) is also invariant under the diffeomorphism
 transformations. 

If a constraint
\be B_{ab} = \e_{abcd} \, e^c \wedge e^d  \,, \ee
is imposed in the action (\ref{BFCG}), one obtains a theory  which is equivalent to the Einstein-Cartan formulation of GR  
\be S_{EC} = \int_M \e^{abcd}\, e_a \wedge e_b \wedge R_{cd} \,.\ee
More precisely
\be
S_{EC} \cong \int_M \left[B_{ab}\wedge R^{ab}+e_{a}\wedge G^{a}-\phi^{ab}\wedge\left(B_{ab}-\epsilon_{abcd}e^c\wedge e^d\right)
\right]\,,\label{ctgr}\ee
see \cite{2pgr}.

\section{Canonical analysis of BFCG theory}

The canonical analysis of the BFCG action can be performed by using the Dirac procedure (DP). This is generally a laborious 
procedure, since it requires the introduction of the canonically conjugate momenta for every variable in the action (\ref{2p}) 
and then executing the DP steps, see \cite{cbf} in the case of a BF theory. However, in certain cases one can obtain a desired 
result in an easier fashion. Namely, given an action for variables $Q$
\be S = \int_I L(Q,\dot Q) \, dt \,,\label{gea}\ee
where $\dot Q = dQ/dt$, then the end-result of the Dirac procedure will be described by the action
\be S_D = \int_I dt \left[ P\,\dot Q - H_0(P,Q) - \l^a \,G_a (P,Q) -\m^\a \,\th_\a (P,Q)\right] \,, \label{da} \ee 
where $P$ are the canonically conjugate momenta for the coordinates $Q$, $G_a$ are the First Class (FC) constraints, 
$\th_\a$ are the Second Class (SC) constraints and $\l$ and $\m$ are the corresponding Lagrange multipliers\footnote{Here $Q$ 
denotes both the set of the coordinates and the corresponding vector. Hence $P\dot Q$ denotes the scalar product of vectors $P$
 and $\dot Q$.}. 

The FC constraints will satisfy
\be \{ G_a , G_b \}_D = f_{ab}^{\,\,\,\, c} (P,Q)\, G_c \,,\ee
and
\be \{G_a ,H_0 \}_D = h_a^{\,b} (P,Q) \,G_b \,,\ee
where
\be \{A,B\}_D = \{A,B\} - \{A,\th_\a\}\D^{\a\b}\{\th_\b , B\} \,,\ee
is the Dirac bracket. $\D^{\a\b}$ is the inverse matrix of $\{\th_\a,\th_\b \}$ and the Poisson Bracket (PB) is defined as
\be \{A,B\} = \frac{\pa A}{\pa Q}\frac{\pa B }{ \pa P }-\frac{\pa A }{ \pa P }\frac{\pa B}{\pa Q }\,.\ee

In particular, if one can write the action (\ref{gea}) in the form
\be S = \int_I dt \left[ p\,\dot q - \l^k \,G_k (p,q) \right] \,,  \label{gfa}\ee 
where $p \cup q \cup \l = Q$ and
\be \{ G_k , G_l \}^* = f_{kl}^{\,\,\,\, m} (p,q)\, G_m \,,\ee
where $\{,\}^*$ is the $(p,q)$ Poisson bracket,
then from (\ref{da}) it follows that (\ref{gfa}) is a gauge-fixed form of $S_D$ where the second-class constraints have been eliminated and some of the phase-space coordinates have been set to zero. Hence the remaining FC constraints are given by $G_k$ and $H_0 \equiv 0$.
	
This approach works in the BFCG case, which can be seen by spliting all the fields into the temporal and the spatial 
comonents via the coordinate splitting 
\be x^\m =(x^0, x^i)=(t,\vec x)\,,\ee
which corresponds to spacetime manifold $M$ having the topology $\S \times I$, where $\S$ is a spatial 3-manifold.

We can then decompose the tensor fields from the action (\ref{2p}) as
\be X_{ \m \cdots} \, Y^{ \m \cdots}=  X_{ 0 \cdots}\,Y^{ 0 \cdots} + X_{ i \cdots}\,Y^{ i \cdots}\,.\ee  
For example    
\be \e^{\m\n\r\s}\,B^{ab}_{\m\n}\,R^{cd}_{\r\s} = 2 \e^{ijk} ( B^{ab}_{0i} R^{cd}_{jk} + B_{ij} R^{cd}_{0k} ) \,, \ee
where
\be  R^{ab}_{\m\n} = \pa_{[\m} \o^{ab}_{\n]} + \o^{ac}_{[\m|} \o^b_{c|\n]} \,. \ee
Similarly
\be \e^{\m\n\r\s}\,e^a_{\m}\,G_{a\, \n\r\s} = \e^{ijk} ( 3 e^a_
{i}\,G_{a\,0jk} -  e^a_{0}\,G_{a\, ijk} ) \,,\ee
where
\be G^a_{\m\n\r} = \pa_{[\m} \b^a_{\n\r]} + \o^{ab}_{[\m|} \b_{b\,|\n\r]}  \,.\ee

The Lagrangian density $\cal L$ of the BFCG action (\ref{2p}) can be written as
		\begin{eqnarray}
			\mathcal{L}&=&\lc{\mu\nu\rho
\sigma}\left(\frac{1}{4}\U{B}{ab}{\mu\nu}\,\vD{R}{ab\rho\sigma}+									
\frac{1}{6}\,\U{e}{a}{\mu}\,\vD{G}{a\nu\rho\sigma}\right)\,,
		\end{eqnarray}
so that
		\begin{eqnarray}
			\mathcal{L}&=&\MD{\p}{ab}{k} \U{\dot\omega}{ab}{k}+\MD{\P}{a}{ij}	\,	
			\U{\dot\beta}{a}{ij}-\mathcal{H}\,,
		\end{eqnarray}		
where
\be \p^{ab\,i} = \frac{1}{2}\,\e^{ijk} B^{ab}_{jk}\,,\quad \P^{a \,ij} = -\frac{1}{2}\,\e^{ijk}\,e^a_k\,, \ee 	
and
\begin{eqnarray}
\mathcal{H}&=&-\left[\frac{1}{2}\,\U{B}{ab}{0i}\e^{ijk}\,{R}_{ab\,jk}+\frac{1}{2}\,\U{e}{a}{0}\e^{ijk}\nabla_i	
				\,{\beta}_{a\,jk}+\right.\cr
				&+&\left. \,\U{\omega}{ab}{0}\left(\nabla_i\,\MD{\pi}{ab}{i}+\,
				\MD{\Pi}{[a}{ij}\,{\beta}_{b]\,ij}	\right)
				+\,{\beta}_{a\,0k}\e^{ijk} \nabla_i\,{e}_{aj}
				\right]\,.\label{2pc}
		\end{eqnarray}
We have discarded a total divergence term in $\cal H$ because
a total divergence vanishes when $\S$ is compact. For non-compact $\S$ we assumed that all fields 
vanish at a spatial infinity. 

The expression  (\ref{2pc}) implies that the constraints  are given by
\begin{eqnarray}
			\mathcal{C}_{ 1 ab}^{\phantom{1ab}i}&\equiv&\frac{1}{2}\e^{ijk}\,{R}_{ab\, jk}=0\,,\cr
			\mathcal{C}^{\phantom 2 a}_{2\,\phantom{a}}&\equiv&\frac{1}{2}\lc{ijk}\nabla_i\beta^{a}_{\phantom a jk}=0\,,\cr
			\mathcal{G}_{1\,ab}&\equiv& \nabla_i\MD{\p}{ab}{i}-\,{\b}_{[a|ij}\,{\P}_{b]}^{ij}=0\,,\cr
			\mathcal{G}^{\phantom{2\,a}k}_{2\,a}&\equiv& \nabla_i\P_a^{\phantom a ik}=\frac{1}{2}\lc{ijk}T^a_{ij}=0 \,,
		\label{cp}\end{eqnarray}
where $T_{ij}$ are the spatial components of the torsion tensor, see (\ref{tor}). The PB algebra of the constraints (\ref{cp}) is given by
		\begin{eqnarray}
		\pb{\mathcal{C}^{\phantom 1 a}_{2}( \vec x)}{\mathcal{G}_{2\,b}^{\phantom{2\,b}i}(\vec y)}= 
-4\mathcal{C}{_{1}}^{a \phantom b i}_{\phantom{a}b}\dir{x}{y}\cr
		\pb{\mathcal{C}^{\phantom 1 a}_{2}(\vec x)}{\mathcal{G}_{1\,cd}(\vec y)}=\kr{a}{[c}\mathcal{C}_{2d]}\dir{x}{y}\cr
	\pb{\mathcal{C}^{\phantom 1 ab\,i}_{1}(\vec x)}{\mathcal{G}_{1\,cd}(\vec y)}=-4\kr{[a}{[c}
           \mathcal{C}_{1\phantom{b]}d]}^{\phantom{1}b]\phantom{b]}i}\dir{x}{y}\cr
	\pb{\mathcal{G}_{1\,ab}(\vec x)}{\mathcal{G}_1^{\phantom 1 cd}(\vec y)}= 4
\kr{[c}{[a}\mathcal{G}_{1\phantom{d]}b]}^{\phantom{3}d]\phantom{b]}}\dir{x}{y}\,,\cr
	\pb{\mathcal{G}_{1\,ab}(\vec x)}{\mathcal{G}_2^{\phantom 1 c}(\vec y)}= 
-\kr{c}{[a}\mathcal{G}_{2\phantom{d]}b]}\dir{x}{y} \,.
	\label{a2p}\end{eqnarray}
Hence the constraints ${\cal C}_k$ and ${\cal G}_k$ are first class and $H_0 \equiv 0$.
 	
\section{$BF$ formulation of Poincar\'e BFCG theory}

Note that the $e\wedge\nabla \b$ term in the BFCG action (\ref{2p}) can be integrated by parts, so that	
 		\begin{eqnarray}
		S &=&\int_M \left(B^{ab}\wedge R_{ab}+e^{a}\wedge \nabla\beta_{a}\right)
  		=\int_M \left[B^{ab}\wedge R_{ab}+e^{a}\wedge \left(d\beta_{a}+\omega_{a}^{\phantom{a}b}\wedge\beta_{b}\right)
                \right]\cr
  		&=&\int_M \left[B^{ab}\wedge R_{ab}+\left(de^{a}+\omega^{ab}\wedge e_{b}\right)\wedge \beta_{a}\right]-\int_M 
d\left(e^a \wedge \beta_a\right)\,. \end{eqnarray}
Hence
\be S \cong \int_M \left(B^{ab}\wedge R_{ab}+T^{a}\wedge \beta_{a}\right) \,,\label{bfp}\ee
where
\be T^a = d e^a + \omega^{a}_{\phantom a c} \wedge e^c \,,\label{tor}\ee
is the torsion.
	
The action (\ref{bfp}) represents a BF-theory action for the Poincar\'e group, which can be seen by introducing a Poincar\'e-group connection
\be A(x) = A^I(x) X_I = \omega^{ab}(x)\, J_{ab} + e^a (x) \, P_a \,,\ee
where $J$ and $P$ satisfy the Poincar\'e Lie algebra
\be [J_{ab},J_{cd}] = \et_{[a|[c}J_{d]|b]}  \,,\quad
 [P_a,J_{bc}] = \et_{a[b}P_{c]}  \,, \quad [P_a,P_b] = 0 \,.\ee

The corresponding curvature  is given by
 	\be
 		F=F^I X_I = \left( d A^I + f^{\quad I}_{JK}\, A^J \wedge A^K \right) X_I \,,
 	\ee
so that
 \be
 		F = R^{ab} J_{ab}+ T^{a} P_a\,.
 	\ee
  	
The action (\ref{bfp}) can be then written in a BF form as
 	\begin{eqnarray}
		 S=\int_M  B^{I}\wedge F_{I} \,,
 	\end{eqnarray}
 where
 	\be
 	 B^{I} = \left(B^{ab}, \beta^a \right)\,,\quad
 		F_{I}= \left(R_{ab}, T_a \right)\,.
 	\ee
 
The  canonical analysis can be performed by using the same method as in the BFCG case. The Lagrangian density can be written as 
 		\begin{eqnarray}
			\mathcal{L}&=&\MD{\p}{ab}{i}\,\U{\dot\omega}{ab}{i}+\MD{p}{a}{i}\,\U{\dot e}{a}{i}-
\tilde{\cal H}\,,
		\end{eqnarray}
where    		
 		\be
			\MD{\p}{ab}{i} = \frac{1}{2}\e^{ijk}{B}_{ab\,jk}\,,\quad
			\MD{p}{a}{i} = \frac{1}{2}\e^{ijk}{\beta}_{a\,jk}\,,
		\ee
and
\begin{eqnarray}
			\tilde{\cal H}&=&-\left[\frac{1}{2}\e^{ijk}\U{B}{ab}{0i}\,{R}_{ab \,jk}+\,\U{e}{a}{0}
\nabla_i \MD{p}{a}{i}+\right.\cr
			&+&\left. \,\U{\omega}{ab}{0}\left(\nabla_i\,\MD{\p}{ab}{i}-{e}_{[a|i}\,\MD{p}{b]}{i}\right)	
				+\frac{1}{2}\e^{ijk}\,{\beta}_{a\,0i} \,\U{T}{a}{jk} \right]\,.
		\end{eqnarray}

Therefore the constraints are given by 
	\begin{eqnarray}
	\tilde{\cal C}_1^{\phantom 1 abi}&\equiv&\frac{1}{2}\lc{ijk}\U{R}{ab}{jk}=0,\\
	\tilde{\cal C}^{\phantom 2 a\,i}_{2}&\equiv&\frac{1}{2}\lc{ijk} T^a_{\phantom a jk}=0\\
	\tilde{\cal G}_{1\,ab}&\equiv&\nabla_i\,\MD{\p}{ab}{i}-{e}_{[a|i}\,\MD{p}{b]}{i}=0 \,,\\
	\tilde{\cal G}_{2\,a}&\equiv&\nabla_ip_a^{\phantom a i}=0 \,.
	\end{eqnarray}
The PB algebra of these constraints is given by 
\begin{eqnarray}
		\pb{\tilde{\cal C}^{\phantom 1 a}_{2}(\vec x)}{\tilde{\cal G}_{2\,b}^{\phantom{2\,b}i}(\vec y)}= 
-4\tilde{\cal C}{_{1}}^{a \phantom b i}_{\phantom{a}b}\dir{x}{y}\cr
		\pb{\tilde{\cal C}^{\phantom 1 a}_{2}(\vec x)}{\mathcal{G}_{1\,cd}(\vec y)}=\kr{a}{[c}\tilde{\cal C}_{2d]}
\dir{x}{y}\cr
	\pb{\tilde{\cal C}^{\phantom 1 ab\,i}_{1}(\vec x)}{\tilde{\cal G}_{1\,cd}(\vec y)}=-4\kr{[a}{[c}
           \tilde{\cal C}_{1\phantom{b]}d]}^{\phantom{1}b]\phantom{b]}i}\dir{x}{y}\cr
	\pb{\tilde{\cal G}_{1\,ab}(\vec x)}{\tilde{\cal G}_1^{\phantom 1 cd}(\vec y)}= 
4\kr{[c}{[a}\tilde{\cal G}_{1\phantom{d]}b]}^{\phantom{3}d]\phantom{b]}}\dir{x}{y}\,,\cr
	\pb{\tilde{\cal G}_{1\,ab}(\vec x)}{\tilde{\cal G}_2^{\phantom 1 c}(\vec y)}= 
-\kr{c}{[a}\tilde{\cal G}_{2\phantom{d]}b]}\dir{x}{y} \,.
	\label{apbf}\end{eqnarray}
Hence the constraints $\tilde{\cal C}_k$ and $\tilde{\cal G}_k$ are first class and $H_0 \equiv 0$.

Note that the BF constraint algebra (\ref{apbf}) is the same as the BFCG constraint algebra (\ref{a2p}). This is because there is a canonical transformation which relates the canonical pairs $(\beta,\P)$ and $(e,p)$. It is given by
	\be\label{ctr}
	\U{\beta}{a}{ij} = \varepsilon_{ijk}p^{ak}\,,\quad
	\MD{\P}{a}{ij} = -\lc{ijk}e_{ak} \,,
	\ee
so that $\tilde{\cal C}_k = {\cal C}_k$ and $\tilde{\cal G}_k = {\cal G}_k$. Hence (\ref{ctr}) transforms the Poincar\'e BFCG theory into the  BF theory for the Poincar\'e group. 
	
\section{Canonical quantization}

Given a set of canonical variables $\{(p_k , q_k) | \, k\in K\}$, one can define a quantization based on a representation of the corresponding Heisenberg algebra in the Hilbert space ${\cal H}_0 = L_2 \left({\bf R}^{|K|}\right)$ such that
\be \hat p_k \,\Psi (q) = i\,{\pa\Psi (q)\over\pa q^k} \,,\quad \hat q_k \,\Psi (q) = q_k \,\Psi (q)	\,.\label{qb}\ee
We will refer to the representation (\ref{qb}) as the quantization in the $q$ basis.

The results of the previous section imply that the canonical quantization  of the Poincar\'e BFCG theory in the 2-connection basis $(\o,\b)$, can be related to 
the canonical quantization of the Poncare BF theory in the $(\o,e)$ basis. Since $\b$ is canonically conjugate to $e$, by performing a functional Fourirer transform, we obtain
	\be \Psi (\o , \b) = \int {\cal D}e \,\Phi (\o,e)\exp\left(i\int_\S \b^a \wedge e_a\right) \,. \label{ft}\ee
	
On the other hand, $\Phi(\o,e)\equiv \Phi(A)$ is a solution of a quantum version of the Poincare BF constraints. For any 
BF theory, the canonical pair $(A_i^I,E^i_I )$ can be represented by the operators
\be \hat E^i_I (x) \,\Phi (A) = i \frac{\d \Phi}{\d A_i^I(x)} \,,\quad \hat A_i^I (x)\, \Phi (A) = A_i^I (x)\, \Phi (A) \,,\ee
so that the Gauss constraint
\be \hat G_I \,\Phi (A) = \pa_i \left(\frac{\d \Phi}{\d A_i^I (x)}\right) + f_{IJ}^{\,\,\,\,K}  A_i^J(x) \,
\frac{\d \Phi}{\d A_i^K (x)} = 0 \,,\ee
is equivalent to 
\be \Phi(A) = \Phi (\tilde A)\, \ee
where $\tilde A = A + d\l + [A, \l]$ is the infinitesimal gauge-transform of $A$. This implies that $\Phi(A)$ must be a 
gauge-invariant functional, while the vanishing curvature constraint
\be F(A(x))\Phi (A) = 0 \ee
implies
\be \Phi(A) = \prod_x \d(F_x) \phi (A)\,, \ee
i.e. $\Phi(A)$ has a non-zero support on flat connections.

Consequently any gauge-invariant functional of flat Poincar\'e connections on $\S$, $\Phi(\o_0,e_0)$, is a solution. 
The space of $\Phi(\o_0,e_0)$, which we denote as ${\cal H}_0$, is the space of functions on the moduli space of flat 
connections on $\S$ for the Poincare group $ISO(1,3)$, which we denote as $MS(ISO(3,1))$. It is easy to see that
	\be MS(ISO(3,1)) = VB[MS(SO(3,1))]\,, \label{fcc}\ee
where VB is the vector bundle such that the fiber at a point $\o_0$ of $MS(SO(3,1))$ is the solution space of the 
vanishing torsion $de_0 + \o_0 \wedge e_0 =0$.

In ${\cal H}_0$ we can introduce a basis of spin-network wavefunctions. Let $A$ be a connection for a Lie group $G$ on $\S$, 
and let $\g$ be a graph in $\S$. Given the irreps $\L_l$ 
of $G$ associated to the edges of $\g$ and the corresponding intertwiners $\i_v$ associated to the vertices of $\g$, 
one can construct the spin-network wavefunctions
\be W_{\hat\g} (A) = Tr\,\left(\prod_{v\in\g} C^{(\i_v)} \prod_{l\in\g} D^{(\L_l)}(A) \right) \equiv \langle A | \hat\g \rangle \,,
\label{snwf}\ee
where $D^{(\L_l)}(A)$ is the holonomy for the line-segment $l$, $C^{(\i)}$ are the intertwiner coefficients and $\hat\g =(\g,\L,\i)$ denotes a spin network associated to a graph $\g$.

Note that when $A$ is a flat connection, than (\ref{snwf}) is invariant under a homotopy of the graph $\g$, so that
we can label the spin-network wavefunctions by combinatorial (abstract) graphs $\g$.

In the case of a non-compact group there is a technical difficulty when constructing the spin-network wavefunctions. Namely, 
if one uses the unitary irreps (UIR), these are infinite-dimensional, and one has to insure that the trace in (\ref{snwf}) is 
convergent. In the Poincare group case, we will consider the massive UIRs, which are labelled by a pair $(M,j)$, where $M > 0$ 
is the mass and
$j\in {\bf Z}_+/2$ is an $SU(2)$ spin. In this case
\be D^{(M,j)}_{q,m';p, m}(\o,a) = e^{i (\L_\o p) \cdot a} \,D^{(j)}_{m'm}(W(\o,p))\, \d^3 ({\vec q} -\vec{\L_\o p})\,, \ee
where $p=(p_0,\vec p) = (\sqrt{(\vec p)^2 + M^2} ,\vec p )$, $D^{(j)}$ is a spin-$j$ rotation matrix and $W(\o,p)$ is the 
Wigner rotation, see \cite{hl}.

By requiring that $W_{\hat\g}(A)$ form a basis in ${\cal H}_0$, we obtain
\be |\Psi \rangle = \int DA \,|A\rangle \langle A|\Psi \rangle = \sum_{\hat\g} |\hat\g\rangle \langle \hat\g | \Psi \rangle \ee
and
\be\langle \hat\g | \Psi \rangle = \int DA \,\langle \hat\g |A\rangle \langle A|\Psi \rangle = \int DA\, W^*_{\hat\g} (A)\, 
\Psi (A) \,. \ee
The last formula is known as the loop transform.

Since we are dealing with a Lie 2-group, one would like to generalize the spin-network wavefunctions for the case of a 2-connection $(\o,\b)$. The categorical nature of a 2-group implies that one can associate 2-group representations to a 2-complex.
Namely, if $(\o,\b)$ is a 2-connection for a Lie 2-group $(G,H)$ on $\S$, then given a 2-complex $\G$ in $\S$, one can associate 
the 2-group representations $L_f$ to the faces $f$ of $\G$. The corresponding 1-intertwiners $\L_l$ can be associated to the edges
of $\G$, while the corresponding 2-intertwiners $\i_v$ can be associated to the vertices of $\G$. Hence we obtain a spin foam
$\hat\G = (\G,L,\L,\i)$.

For example, in the 2-Poincare group case, there is a class of representations labelled by a positive number $L$, see \cite{bbfw}.
The intertwiners for 3 such representations, $L_1,L_2,L_3$, are labeled by integers $m$ if $L_k$ satisfy the triangle 
inequalities strongly. The $m$'s label the irreps of an $SO(2)$ group, which leaves the triangle $(L_1,L_2,L_3)$, embedded in 
${\bf R}^4$, invariant. The 2-intertwiners for the $m$'s are trivial and $L_k$ in this case can be identified with the 
edge-lengths of a triangle, see \cite{2pgr}. If $L_k$ are collinear, i.e. $L_1 = L_2 + L_3$, the invariance group is $SO(3)$ and 
the corresponding intertwiners are the $SU(2)$ spins $j$ while the 2-intertwiners are the $SU(2)$ intertwiners. In this case the 
$L_k$ look like particle masses, but then it is not clear what would be the geometrical interpretation of these masses.

A spin-foam wavefunction should be an appropriate generalization of the spin-network wavefunction (\ref{snwf}) such that the spin-foam wavefunction includes the surface holonomies associated with the spin-foam faces $f$. Let us embed $\G$ into a triangulation of the spatial manifold and let $\o$ and $\b$ be piece-wise constant in the appropriate cells of the triangulation. If $g_l = \exp (\o_l J)$ and $h_f = \exp (\b_f P)$, then the formula for the surface holonomy $h_p$ for the surface of a polyhedron $p$, is given by
\be h_p =\prod_{f\in \partial p}g_{l(f)}\triangleright h_f \,,\ee
where $g_{l(f)}$ can be calculated by representing the $\pa p$ surface as a composition 
of 2-morphisms $(g_l,h_f)$ from some 1-morphism $g_{l'}$ ($l' \in p$) to itself, see \cite{gpp} for the case of a tetrahedron.

Hence we expect that
\be W_{\hat\G} (\o,\b) = Tr\left(\prod_{v\in\G} C^{(\i_v )}\prod_{l\in\G}D^{(\L_{l})}(\o) \prod_{f\in\G} D^{(L_f)}(\o,\b) \right)\,,\label{sfwf}\ee
where 
\be D^{(L_f)}(\o,\b) = D^{(L_f)}(g_{l(f)} \triangleright h_f )\,.\label{dhm}\ee
In the Poincare 2-group case, the representation matrix (\ref{dhm}) is of the type $1\times 1$, because $H$ is an abelian group. The analysis in \cite{2pgr} suggests that
\be D^{(L_f)}(\o,\b) = \exp\left(i\vec L_f \cdot g(\o_l)\vec\b_f \right) \,,\ee
where $\vec L_f$ is a 4-vector satisfying $L_f^2 =  \vec L_f \cdot\vec L_f = \eta_{ab}\,L_f^a L_f^b$ and $\eta$ is a flat Minkowski metric.

A related problem is that it
is not known what is the 2-group analog of the Peter-Weyl theorem
\be \f(g) =\sum_\L \sum_{\a_\L , \b_\l} \tilde \f_\L^{\,\a_\L \b_\l}\, D^{(\L)}_{\a_\L \b_\l}(g) = \sum_\L \langle \tilde \f_\L
 \,, D^{(\L)}(g)\rangle \,,\ee
where $\f$ is a function on a Lie group $G$ and
\be \tilde \f_\L^{\,\a_\L \b_\l} = \int_G dg \,\bar D^{(\L)}_{\a_\L \b_\l}(g)\,\f(g) \,. \ee

Note that in the case of the Poincar\'e 2-group, the relation (\ref{ft}) can give some clues. Let us consider again piece-wise constant fields on a triangulated manifold. The Poincar\'e group holonomy for an edge $\e$ is given by $g'_\e = \exp(\o_\e J + e_\e P)$, so that a function $\f(g'_\e)=\F(\o_\e,e_\e)$ can be expanded by using the generalization of the PW theorem for the Poincar\'e group
\be \F (\o_\e ,e_\e ) = \int_0^\infty dM \, \sum_j \langle {\tilde \F}_{M,j} \,,\, D^{(M,j)}(\o_\e ,e_\e)\rangle \,.\ee
Consequently
\bea  \Psi (\o_\e ,\b_f ) &=& \int_{{\bf R}^4} d^4 e_l \,\m(e_\e)\,e^{i\vec\b_f \cdot\vec e_{\e}}\, \F(\o_\e,e_\e)\cr
&=&
\int_0^\infty dM \sum_j \langle \tilde \F_{M,j} \,,\,
\int_{{\bf R}^4} d^4 e_\e \,\m(e_\e)\,e^{i\vec\b_f \cdot \vec e_{\e}}\, D^{(M,j)}(\o_\e ,e_\e )\rangle \cr 
&=& \int_0^\infty dM \sum_j \langle \tilde \F_{M,j} \,,\, \tilde D^{(M,j)}(\o_\e ,\b_f )\rangle \,,\label{ppw}\eea
where $\m$ is some appropriatelly chosen measure and $f$ is the face dual to an edge $\e$.

If $\G$ is a tetrahedron, then by comparing (\ref{ppw}) to (\ref{sfwf}) one concludes that $\L_\e = j_\e$ and that there should be a relationship between an $L_\D$ and the three $M_{f}$ for the dual faces for the edges of a triangle $\D$. Furthermore, there should be a relationship between the functions $D^{(j_\e)}(g_\e) D^{(L_\D)}(g_{\e'},h_\D)$ on $\G$ and the functions $\tilde D^{(M_f , j_\e)}(g'_\e)$ on $\G$.

\section{Conclusions}
	
We have found a canonical formulation of the BFCG action for the Poincar\'e 2-group where the phase-space variables are the 
2-connection $(\o_i^{ab},\b_{ij}^a)$ on a 3-manifold $\S$ and its canonically conjugate pair $(\p^i_{ab},\P^{ij}_a)$. This 
canonical formulation is suitable for the canonical quantization where the physical Hilbert space is spanned by the spin-foam 
states, which are the categorical generalization of the spin-network states from LQG. By using the fact that the BFCG action for 
the 2-Poincar\'e group is equivalent to the BF action for the
Poincar\'e group, we obtain a canonical transformation which relates the two canonical formulations. In the BF canonical 
formulation, the basic variable is the Poincare connection $(\o_i^{ab},e_{i}^a)$ on $\S$ and its canonically conjugate pair 
$(\p^i_{ab},p^{i}_a)$, and the corresponding cannonical transformation is given by (\ref{ctr}).

There is a mathematical difficulty when trying to construct the spin-foam basis, that comes from the lack of knowledge of what 
is the exact form of the Peter-Weyl theorem for 2-groups. However, in the Poincar\'e 2-group case, we can use the relation to 
the Poincar\'e BF theory, which gives important clues how to construct the spin-foam wavefunctions. We beleive that those clues 
will be sufficient to complete the spin-foam basis construction.

On the other hand, one can quantize the theory in the BF formulation, and in this case the physical Hilbert space is given by 
the space of square-integrable functions on the moduli space of flat connections. One can proceed further, and introduce the 
spin-network basis, by constructing the spin-network wave functions for the Poincare group. An interesting problem will be 
to investigate the relation between the spin-network basis and the spin-foam basis.

As far as the canonical quantization of GR in the spin-foam basis is concerned, this requires a canonical formulation of the constrained BFCG theory based on the 2-connection variables $(\o,\b)$ and their momenta $(\p,\P)$. However, the structure of the GR constraints is such that the short-cut procedure based on the space-time decomposition of the fields in the action does not work, and one has to perform the full Dirac procedure. Given the corresponding action $S_D$, one has to eliminate the second-class constraints by using a gauge-fixing procedure in order to obtain the action (\ref{gfa}) for an appropriate subset of the BFCG variables and their conjugate momenta.

We expect that the reduced variables can be chosen as $(\tilde\o_{i}^{\a},\tilde\p^{i}_{\a})$ and $(\tilde\b_{ij}^\a,\tilde\P^{ij}_\a)$, where $\a =1,2,3$ and $\b_{ij}^a = (\b_{ij}^0 ,\tilde\b_{ij}^\a)$ while $\o_i^{ab} = (\o_i^{0\a} ,\e^{\a\b\g}\tilde\o_{i\g})$. Hence the gauge choice will be to set $\b^0_i$ and $\o_{ij}^{0\a}$ components to zero. Then we can consider $(\tilde\o_i^\a,\tilde\b_{ij}^\a)$ as a 2-connection for the three-dimensional Euclidean 2-group $(SO(3),{\bf R}^3)$. The dual variable $\tilde e_i^\a = \e_{ijk} \tilde\P^{jk\a}$ can be considered as a triad, so that the FC constraints for the $(\tilde e,\tilde\o,\tilde p,\tilde\p)$ variables, where $\tilde p^i_\a = \e^{ijk}\tilde\b_{jk\a}$, should give the triad canonical formulation of GR when the connection $\tilde\o$ is eliminated by the torsion constraint $T_{ij}^\a = \partial_{[i} \tilde e_{j]}^\a + \e^{\a\b\g}\tilde\o_{[i|\b} \tilde e_{|j]\g} = 0$. This implies that the FC constraints for the variables $(\tilde e,\tilde\o,\tilde p,\tilde\p)$ variables should be
\be {\cal H}(\tilde e,\tilde\o,\tilde p,\tilde\p)=0\,,\, D_i(\tilde e,\tilde\o,\tilde p,\tilde\p) = 0 \,,\, G_\a (\tilde e,\tilde\o,\tilde p,\tilde\p) = 0 \,,\, T_{ij}^\a (\tilde e,\tilde\o) =0 \,,  \ee
where $\cal H$ is the Hamiltonian constraint, $D_i$ is the 3-diffeomorphism constraint and $G_\a$ is the Gauss constraint for the $SO(3)$ group. Although the form of $G_\a$ and $D_i$ can be guessed, the form of $\cal H$ is not obvious, and requires a further work. By making the canonical transformation $(\tilde e,\tilde p)\to (\tilde\b,\tilde\P)$ one would obtain the FC constraints for the 2-connection variables $(\tilde\o,\tilde\b,\tilde\p,\tilde\P)$.

In this way one would generalize the LQG spin-network basis to a spin-foam basis, and a hope is that the corresponding Hamiltonian constraint may be simpler to solve. The definite advantage over the LQG formalism is that 
one can construct a wavefunction which is a function of the triads $\tilde e$ and the connection $\tilde\o$, so that
it will be easier to perform the semi-classical analysis. 

\bigskip

\noindent{\bf Acknowledgements}

\bigskip
\noindent We would like to thank M. Vojinovi\'c for discussions. A. Mikovi\'c was partially supported by the FCT grants PEst-OE/MAT/UI0208/2011 and EXCL/MAT-GEO/0222/2012, while M. Oliveira was supported by the FCT PhD grant 
SFRH/BD/79285/2011.

 \end{document}